\documentstyle[12pt]{article}

\begin{document}

\thispagestyle{empty} \renewcommand{\thefootnote}{\fnsymbol{footnote}}

\begin{center}
\vspace{1truecm} {\large {\bf Light-Cone Formulation of Super D2-Brane}}

\vspace{1cm} {\bf R. Manvelyan$^{1}$ \footnote{E-mail:
manvel@physik.uni-kl.de} \footnote{Alexander von Humboldt Fellow,\\
\,\,\,\,\,\,\,\,\,\, On leave from Yerevan Physics Istitute}, A. Melikyan$^{2}$ 
\footnote{E-mail: arsen@moon.yerphi.am,}}

{\bf R.Mkrtchyan$^{2}$ \footnote{E-mail: mrl@amsun.yerphi.am}, H.J.W.
M\"{u}ller--Kirsten$^{1}$ \footnote{E-mail: mueller1@physik.uni-kl.de}}%
\vspace{1cm}

$^{1}${\it Department of Physics, University of Kaiserslautern,}

{\it P.\ O.\ Box 3049, D 67653 Kaiserslautern, Germany} \vspace{1cm}

$^{2}${\it Theoretical Physics Department,} {\it Yerevan Physics Institute,}

{\it Alikhanian Br. St.2, Yerevan, 375036 Armenia }
\end{center}

\vspace{1cm}

\renewcommand{\thefootnote}{\arabic{footnote}} \setcounter{footnote}0
%\newpage

\renewcommand{\thefootnote}{\arabic{footnote}} \setcounter{footnote}0
\begin{abstract}
The light--cone Hamiltonian approach is applied to the super D2-brane, and
the equivalent area--preserving and U(1) gauge--invariant effective
Lagrangian, which is quadratic in the U(1) gauge field, is derived. The
latter is recognised to be that of the three--dimensional U(1) gauge theory,
interacting with matter supermultiplets, in a special external induced
supergravity metric and the gravitino field, depending on matter fields. The
duality between this theory and 11d supermembrane theory is demonstrated in
the light--cone gauge.
\end{abstract}

{\smallskip \pagebreak }

\section{Introduction}

In recent years significant progress was made in the theory of
higher--dimensio-nal extended supersymmetric objects. Extended objects play
a central role in our understanding of non-perturbative aspects of
superstring and supergravity theories. Moreover, the two-brane plays a
crucial role in M-theory\cite{MTH}. M-theory is intrinsically
eleven-dimensional, and includes, in the spectrum of excitations,
eleven-dimensional supermembrane theory. Besides the ``usual'' p-branes, with 
actions, described by scalars (brane coordinates in space-time) and fermions
(superpartners of scalars), there exists a new class of extended objects,
the so-called D-branes \cite{Pol}. The D-branes contain in their spectrum a
vector field (or, in the case of the eleven-dimensional 5-brane, a second
rank self-dual tensor field). The main feature of the D-brane is that open
superstrings can end on it. This means that open strings with Dirichlet
boundary conditions describe the dynamics of the corresponding D-brane \cite
{Pol} connected by duality transformations with ordinary closed string
backgrounds. The incorporation of D-branes in the superstring picture leads
to a deeper understanding of the theory of solitonic states in
non-perturbative string theory and of various aspects of string/M-theory
dualities.

In ref. \cite{MMM} we investigated the light-cone gauge for the bosonic part
of the action of the D2-brane. The main result of that paper is that the
corresponding gauged light-cone action for the Dirac-Born-Infeld Lagrangian
can be rewritten as a three-dimensional Maxwell theory with matter field in
the specific curved induced metric: 
\begin{equation}
{G}_{\mu \nu }=\left( 
\begin{array}{ll}
-g+\xi ^{i}\left( \omega \right) g_{ij}\xi ^{j}\left( \omega \right) & \xi
^{k}\left( \omega \right) g_{kj} \\ 
\xi ^{k}\left( \omega \right) g_{ki} & g_{ij}
\end{array}
\right)  \label{Met}
\end{equation}
where $\omega (\tau ,\sigma _{i})$ is the area-preserving gauge field, $\mu
,\nu =0,1,2$, and 
\begin{eqnarray*}
g_{ij} &=&\partial _{i}X^{a}\partial _{j}X^{a},\,\,\,\,\xi ^{i}\left( \omega
\right) =\varepsilon ^{ki}\partial _{k}\omega \\
g &=&\det g_{ij},\,\,\,a=1,..8;\,\,\,\,i,j=1,2
\end{eqnarray*}

The main goal of the present paper is to extend our previous investigation
to the supersymmetric case. This leads to the interesting three dimensional
structure of the effective light-cone action for the super D2-brane with the
covariant interaction of abelian gauge and matter supermultiplets with
induced supergravity multiplet formed by (\ref{Met}) and the induced
gravitino field $\psi _{\mu }$. This action is closely related to extended
supergravity in d=3, and to ordinary eleven-dimensional supermembrane theory
in the light-cone gauge \cite{deWitt}. The latter allows the duality
transformation to be established between these two extended objects\footnote{%
The connection between the D2-brane in 10d and the membrane in 11d was first
observed by M. J. Duff and J. X. Lu \cite{Duff}. Schmidhuber \cite{Sch} and
Townsend \cite{THS} established this connection in both directions.} in the
light-cone gauge in terms of our induced metric (\ref{Met}). In this article
we prove that the effective action obtained from the Hamiltonian formulation
for the super D2-brane in the light-cone gauge, exactly coincides with the
action obtained from the supermembrane action in d=11 after duality
transformation with our induced metric.

\section{Hamiltonian formulation of super D2-brane and gauged Lagrangian}

The light-cone formulation of the super-membrane obtained in \cite{deWitt}
is closely connected with the matrix model representation of M-theory \cite
{Banks}. The corresponding area-preserving Lagrangian, from which we can
obtain the matrix model by replacing Lie brackets by commutators, is \cite
{deWitt}:

\begin{eqnarray}
L_{M} &=&\frac{1}{2}(D_{0}X^{\dot{a}})^{2}-\frac{1}{4}\left\{ X^{\dot{a}},X^{%
\dot{b}}\right\} \left\{ X^{\dot{a}},X^{\dot{b}}\right\}  \label{MT} \\
&&-iS^{T}D_{0}S-iS^{T}\gamma ^{\dot{a}}\left\{ X^{\dot{a}},S\right\} 
\nonumber
\end{eqnarray}
where $\dot{a},\dot{b}=1,2,...9$, and $S$ is the $SO(9)$ spinor coordinate,
and $D_{0}=\partial _{0}+\{\omega ,...\}$ is a covariant area-preserving
derivative with gauge field $\omega (\tau ,\sigma _{1},\sigma _{2})$ and Lie
bracket 
\begin{equation}
\{X,Y\}=\varepsilon ^{ij}\partial _{i}X\partial
_{j}Y\,,\,\,\,\,\,\,\,\,\,\,\,\,i,j,..=1,2.  \label{Leebra}
\end{equation}

The Lagrangian (\ref{MT}) can be interpreted as that of a 10-dimensional
super-Yang-Mills theory (if we start from an 11-dimensional target space for
the membrane) reduced to one dimension.

Here we investigate the light-cone formulation of the 10-dimensional super
D-membrane described by the following supersymmetric Dirac-Born-Infeld (DBI)
Lagrangian \cite{Schwarz},\cite{Sch} : 
\begin{equation}
S=-\int d^{3}\sigma \sqrt{-{\rm det}\,(\Sigma _{\mu \nu }+{\cal F}_{\mu \nu
})}-\int (C_{3}+C_{1}\wedge {\cal F}).  \label{DBI}
\end{equation}
Here 
\begin{eqnarray*}
\Sigma _{\mu \nu } &=&\Pi _{\mu }^{M}\Pi _{\nu }^{M},\,\,\,\,\,\Pi _{\mu
}^{M}=\partial _{\mu }X^{M}-i\bar{\theta}\Gamma ^{M}\partial _{\mu }\theta ,
\\
{\cal F}_{\mu \nu } &=&F_{\mu \nu }-b_{\mu \nu }=\partial _{\mu }A_{\nu }-i%
\bar{\theta}\Gamma _{11}\Gamma ^{M}\partial _{\mu }\theta \left( \partial
_{\nu }X^{M}-i{\frac{1}{2}}\bar{\theta}\Gamma ^{M}\partial _{\nu }\theta
\right) -(\mu \leftrightarrow \nu ), \\
M,N &=&0,1,..9;\,\,\,\,\,\mu ,\nu =0,1,2
\end{eqnarray*}
and RR-forms $C_{3}$ and $C_{1}$ are determined by the condition \cite
{Schwarz} 
\begin{equation}
d(C_{3}+C_{1}\wedge {\cal F})=d\bar{\theta}(\frac{1}{2}\psi ^{2}+{\cal F}%
\Gamma _{11})d\theta ,  \label{DBI1}
\end{equation}
with $\psi \equiv \Gamma ^{M}\Pi ^{M}$.

We shall construct the analogue of the area-preserving supermembrane action (%
\ref{MT}) for the DBI case,which will be the supersymmetrization of our
previous result \cite{MMM}.

We start with the Hamiltonian formulation of action $\left( \ref{DBI}\right) 
$ following the scheme of \cite{deWitt} and \cite{MMM} after first imposing
the light-cone gauge condition: 
\begin{eqnarray}
X^{+}(\tau ,\sigma _{i}) &=&X^{+}(0,0)+\tau
,\,\,\,\,\,\,\,\,\,\,\,\,\,\,\,\,\,X^{\pm }=\sqrt{\frac{1}{2}}\left(
X^{10}\pm X^{0}\right)  \label{Gaug} \\
\Gamma ^{+}\theta &=&0,\,\,\,\,\,\,\,\,\,\,\,\,\,\,\,\Gamma ^{\pm }=\sqrt{%
\frac{1}{2}}\left( \Gamma ^{10}\pm \Gamma ^{0}\right).  \label{Gaugf}
\end{eqnarray}
After some tedious but straightforward calculations we can write down the
corresponding final Hamiltonian density and the residual area preserving and
Gauss-law (corresponding to $U\left( 1\right) \,$ gauge invariance)
constraints: 
\begin{eqnarray}
H &=&\frac{1}{2}\left[ P^{a}P^{a}+P_{A}^{i}P_{A}^{j}g_{ij}+ \det
(g_{ij}+F_{ij})\right]  \nonumber \\
&&-P_{A}^{i}\bar{P}_{\theta }\Gamma _{11}\partial _{i}\theta +\bar{P}%
_{\theta }\Gamma ^{a}\left\{ X^{a},\theta \right\} ,  \label{hamf1} \\
\Phi &=&\varepsilon ^{ij}\partial _{i}(P^{a}\partial
_{j}X^{a}+P_{A}^{k}F_{jk}+\bar{P}_{\theta }\partial _{j}\theta )\approx 0,
\label{con1} \\
\Upsilon &=&\partial _{i}P_{A}^{i}\approx 0,  \label{con2} \\
\bar{P}_{\theta } &+&i\bar{\theta }\Gamma ^{-}\approx 0,  \label{con3} \\
a &=&1,2...8;\,\,i,j,k=1,2  \nonumber
\end{eqnarray}
Here $P^{a},P_{A}^{i}\,,\bar{P}_{\theta }$ are momenta corresponding to
fields $X^{a},A_{i}$ and $\theta$. Note that here, as in the bosonic case 
\cite{MMM}, we first fix the gauge $A_{0}=0$ and then put $P^{+}=1$ and
correctly reexpress the $X^{-}$ coordinate through the transverse ones : 
\begin{equation}
\partial _{i}X^{-}=-(P^{a}\partial _{i}X^{a}+P_{A}^{j}F_{ij}+\bar{P}_{\theta
}\partial _{i}\theta )  \label{Res}
\end{equation}
using the additional gauge condition: 
\begin{equation}
\partial _{0}X^{a}\partial _{i}X^{a}+\partial _{i}X^{-}-i\bar{\theta}\Gamma
^{-}\partial _{i}\theta +(\partial _{0}A_{k}-i\bar{\theta}\Gamma^{-}\Gamma
_{11}\partial _{k}\theta )g^{kj}F_{ij}=0  \label{Gaug1}
\end{equation}

Moreover, in the gauge $\left( \ref{Gaug}\right), \left( \ref{Gaugf}\right),
\left( \ref{Gaug1}\right)$ the expressions for the momenta be come very
simple: 
\begin{eqnarray}
P^{a} &=&\partial _{0}X^{a}  \nonumber \\
P_{A}^{i} &=&g^{ij}(\partial _{0}A_{j}-i\bar{\theta}\gamma _{-}\Gamma
_{11}\partial _{j}\theta )  \label{momf} \\
\bar{P}_{\theta } &=&-i\bar{\theta }\Gamma ^{-}  \nonumber
\end{eqnarray}
where 
\begin{equation}
g^{ij}=\frac{\varepsilon ^{ik}\varepsilon ^{jl}g_{kl}}{g}  \label{met11}
\end{equation}

So, finally we can proceed to define the effective gauged Lagrangian
containing fields $X^{M},\theta ,$$A_{i}$ and two gauge fields $\omega
\left( \tau ,\sigma _{i}\right) $ and $Q(\tau ,\sigma _{i})$ with the
following properties:

1. The expressions $\left( \ref{momf}\right) $ and $\left( \ref{hamf1}%
\right) $ have to be derived as standard expressions of the canonical
momenta and Hamiltonian for that Lagrangian in the gauge 
\begin{eqnarray}
\omega \left( \tau ,\sigma _{i}\right) &=&0  \nonumber \\
Q(\tau ,\sigma _{i}) &=&0  \label{gaug2}
\end{eqnarray}

2. The equations of motion for gauge fields $\omega \left( \tau ,\sigma
_{i}\right) $ and $Q(\tau ,\sigma _{i})$ have to coincide (in the gauge $%
\left( \ref{gaug2}\right) $) with corresponding constraints $\left( \ref
{con1}\right) $ and $\left( \ref{con2}\right) $.

3. This Lagrangian has to be gauge invariant under the following gauge
groups:

a) the group of area--preserving diffeomorphisms with generator
corresponding to the constraint $\left( \ref{con1}\right) $,

b) the group of $U\left( 1\right) $ gauge transformations connected with
constraint $\left( \ref{con2}\right) .$

The desired effective Lagrangian has the following form: 
\begin{eqnarray}
L &=&\frac{1}{2}(D_{0}X^{a})^{2}-\frac{1}{4}\left\{ X^{a},X^{b}\right\}
\left\{ X^{a},X^{b}\right\} -i\bar{\theta }\Gamma ^{-}D_{0}\theta -i\bar{%
\theta }\Gamma ^{-}\Gamma ^{a}\left\{ X^{a},\theta \right\}  \nonumber \\
&&+\frac{1}{2}g^{ij}(D_{0}A_{i}-\partial _{i}Q-i\bar{\theta }\Gamma
^{-}\Gamma _{11}\partial _{i}\theta )(D_{0}A_{j}-\partial _{j}Q-i\bar{\theta 
}\Gamma ^{-}\Gamma _{11}\partial _{j}\theta )  \nonumber \\
&&-\frac{1}{2}F_{12}^{2}  \label{lag2}
\end{eqnarray}

where 
\begin{eqnarray}
D_{0}X^{a} &=&\partial _{0}X^{a}-\varepsilon ^{ij}\partial _{i}\omega
\partial _{j}X^{a}  \nonumber \\
&=&\partial _{0}X^{a}-\left\{ \omega ,X^{a}\right\} =\partial
_{0}X^{a}-\pounds _{\xi \left( \omega \right) }X^{a}  \nonumber \\
D_{0}A_{i} &=&D_{0}A_{i}=\partial _{0}A_{r}-\varepsilon ^{kj}\partial
_{i}\partial _{k}\omega A_{j}-\varepsilon ^{kj}\partial _{k}\omega \partial
_{j}A_{i}  \label{covder} \\
&=&\partial _{0}A_{i}-\pounds _{\xi \left( \omega \right) }A_{i}  \nonumber
\\
D_{0}\theta &=&\partial _{0}\theta -\varepsilon ^{ij}\partial _{i}\omega
\partial _{j}\theta  \nonumber
\end{eqnarray}
Here $\pounds _{\xi \left( \omega \right) }$ is the Lie derivative in the
direction of the divergenceless vector field $\xi ^{i}\left( \omega \right)
=\varepsilon ^{ki}\partial _{k}\omega .$ Lagrangian $\left( \ref{lag2}%
\right) $ satisfies all three conditions. The gauge transformations of the
given fields are the following: 
\begin{eqnarray}
\delta _{\varepsilon }X^{a} &=&\left\{ \varepsilon
,X^{a}\right\},\,\,\,\,\,\delta _{\varepsilon }\theta =\left\{ \varepsilon
,\theta \right\}  \nonumber \\
\delta _{\varepsilon }A_{i} &=&\pounds _{\xi \left( \omega \right) }A_{i}, 
\nonumber \\
\,\delta _{\varepsilon }Q &=&\left\{ \varepsilon \,,Q\right\},\,\,\,\,\delta
_{\varepsilon }\omega =\partial _{0}\varepsilon +\left\{ \varepsilon
\,,\omega \right\} ,  \label{gtransf} \\
\delta _{\alpha }X^{a} &=&0,\,\,\,\,\delta _{\alpha }A_{i}=\partial
_{i}\alpha ,\,\,\,\,\,\delta _{\alpha }Q=\partial _{0}\alpha +\left\{ \alpha
\,,\omega \right\} ,  \nonumber \\
\delta _{\alpha }\omega &=&0,\,\,\,\,\,\delta _{\alpha }\theta =0.  \nonumber
\end{eqnarray}
It is easy to see that here as in the bosonic case, the $U\left( 1\right) $
gauge transformations of $Q$ do not commute with the area-preserving ones.
This can be altered by a redefinition of the field $Q$: 
\begin{equation}
A_{0}=Q+\varepsilon ^{ij}\partial _{i}\omega A_{j}  \label{azero}
\end{equation}
Here we introduce the new component $A_{0}$, which unlike $Q$ transforms
(under area-preserving diffeomorphisms) not as a scalar but as the zero
component of a three-dimensional vector field : 
\begin{eqnarray}
\delta _{\alpha }A_{0} &=&\partial _{0}\alpha  \nonumber \\
\delta _{\varepsilon }A_{0} &=&\left\{ \varepsilon \,,A_{0}\right\}
+\partial _{0}\xi ^{i}\left( \varepsilon \right) A_{i}  \label{gtr2}
\end{eqnarray}

Then we can solve the light--cone condition $\left( \ref{Gaugf}\right)$ for
the fermionic coordinate and rewrite our fermionic coordinates and gamma
matrices in terms of $SO(8)$ quantities : 
\begin{eqnarray}
\theta &=&2^{-1/4}\left( 
\begin{array}{c}
S \\ 
0\ 
\end{array}
\right) ,\bar{\theta}\Gamma ^{-}=2^{1/4}\left( 
\begin{array}{ll}
S^{T} & 0
\end{array}
\right) ,\,\,\,\,  \label{Sspinor} \\
\Gamma ^{a} &=&\left( 
\begin{array}{ll}
\gamma ^{a} & 0 \\ 
0 & -\gamma ^{a}
\end{array}
\right) ,\,\,\,\,\Gamma _{11}=\left( 
\begin{array}{ll}
\gamma _{9} & 0 \\ 
0 & -\gamma _{9}
\end{array}
\right).  \label{Gamma}
\end{eqnarray}

With this the Lagrangian $\left( \ref{lag2}\right) $ can be rewritten in the
following form: 
\begin{eqnarray}
L &=&\frac{1}{2}(D_{0}X^{a})^{2}-\frac{1}{4}\left\{ X^{a},X^{b}\right\}
\left\{ X^{a},X^{b}\right\}   \nonumber \\
&&-iS^{T}D_{0}S-iS^{T}\gamma ^{a}\left\{ X^{a},S\right\}   \label{lag3} \\
&&+\frac{1}{2}g^{ij}\tilde{F}_{0i}\tilde{F}_{0j}-\frac{1}{2}F_{12}^{2} 
\nonumber
\end{eqnarray}
where 
\begin{eqnarray*}
\tilde{F}_{0i} &=&F_{0i}-\xi ^{k}F_{ki}-iS^{T}\gamma _{9}\partial _{i}S \\
F_{0i} &=&\partial _{0}A_{i}-\partial _{i}A_{0}
\end{eqnarray*}
Unlike the bosonic case, this Lagrangian is supersymmetric with the
following field transformations: 
\begin{eqnarray}
\delta X^{a} &=&i\varepsilon ^{T}\gamma ^{a}S,  \nonumber \\
\delta \omega  &=&i\varepsilon ^{T}S,  \nonumber \\
\delta S^{T} &=&\frac{1}{2}\varepsilon ^{T}D_{0}X^{a}\gamma _{a}+\frac{1}{2}%
\varepsilon ^{T}g^{ij}\tilde{F}_{0i}\gamma _{9}\gamma _{a}\partial _{j}X^{a}+%
\frac{1}{4}\varepsilon ^{T}\left\{ X^{a},X^{b}\right\} \gamma _{ab}-\frac{1}{%
2}\varepsilon ^{T}F_{12}\gamma _{9},\ \   \nonumber \\
\delta A_{i} &=&i\varepsilon ^{T}\gamma _{9}\gamma _{a}S\partial _{i}X^{a},
\label{susy1} \\
\delta A_{0} &=&i\varepsilon ^{T}\gamma _{9}\gamma _{a}S\partial
_{0}X^{a}+i\delta S^{T}\gamma _{9}S  \nonumber
\end{eqnarray}

\section{Induced 3d formulation and duality}

We introduce the three dimensional induced metric $G_{\mu \nu }$ of $\left( 
\ref{Met}\right)$ with the following properties: 
\begin{eqnarray}
G_{\mu \nu } &=&\left( 
\begin{array}{ll}
-g+\xi ^{i}\left( \omega \right) g_{ij}\xi ^{j}\left( \omega \right) & \xi
^{k}\left( \omega \right) g_{kj} \\ 
\xi ^{k}\left( \omega \right) g_{ki} & g_{ij}
\end{array}
\right) ,  \nonumber \\
G^{\mu \nu } &=&\left( 
\begin{array}{ll}
-1/g & \xi ^{j}\left( \omega \right) /g \\ 
\xi ^{i}\left( \omega \right) /g & g^{ij}-\xi ^{i}\left( \omega \right) \xi
^{j}\left( \omega \right) /g
\end{array}
\right) ,  \label{Met2} \\
\,g_{ij} &=&\partial _{i}X^{a}\partial _{j}X^{a},\,g=\det g_{ij}\,,\,\xi
^{i}\left( \omega \right) =\varepsilon ^{ki}\partial _{k}\omega  \nonumber \\
\det G_{\mu \nu } &=&G,\,\,\sqrt{-G}=g,\,\,\,\,\,\,\,\,\sqrt{-G}G^{00}=-1 
\nonumber
\end{eqnarray}
We then construct the induced three dimensional curved Dirac matrices using
the following definitions: 
\begin{eqnarray}
\gamma _{\mu } &=&(\gamma _{0},\gamma _{i}),\,\,\,\,\gamma _{i}=\partial
_{i}X^{a}\gamma ^{a},  \nonumber \\
\gamma _{0} &=&\gamma +\xi ^{i}(\omega )\gamma _{i},\,\,\,\,\gamma =\frac{1}{%
2}\varepsilon ^{ij}\gamma _{i}\gamma _{j}  \label{3dgamma} \\
\gamma ^{\mu } &=&G^{\mu \nu }\gamma _{\nu }=(-\frac{\gamma }{g},\frac{\xi
^{i}(\omega )\gamma }{g}+g^{ij}\gamma _{j})  \nonumber
\end{eqnarray}
These matrices satisfy a three-dimensional Clifford algebra with our induced
metric $\left( \ref{Met2}\right) $: 
\begin{equation}
\{\gamma _{\mu },\gamma _{\nu }\}=2G_{\mu \nu }  \label{kliff}
\end{equation}
Finally we introduce the induced {\it gravitino field:} 
\begin{eqnarray}
\psi _{\mu } &=&(\psi _{0},\psi _{i}),\,\,\,\,\psi _{i}=\partial
_{i}S,\,\,\,\,\psi _{0}=\xi ^{i}(\omega )\psi _{i}+\gamma _{i}\varepsilon
^{ij}\psi _{j}  \nonumber \\
\psi ^{\mu } &=&G^{\mu \nu }\psi _{\nu }=(\psi ^{0},\psi ^{i})  \label{grav}
\\
\psi ^{0} &=&-\frac{\gamma _{i}\varepsilon ^{ij}\psi _{j}}{g},\,\,\,\,\psi
^{i}=g^{ij}\psi _{j}+\frac{\xi ^{i}(\omega )\gamma _{k}\varepsilon ^{kl}\psi
_{l}}{g}  \nonumber
\end{eqnarray}
With this we can rewrite our light--cone action $\left( \ref{lag3}\right) $
in the following final covariant form on our induced metric: 
\begin{eqnarray}
L &=&-\frac{1}{2}\sqrt{-G}G^{\mu \nu }\partial _{\mu }X^{a}\partial _{\nu
}X^{a}+\frac{1}{2}\sqrt{-G}  \nonumber \\
&&+i\sqrt{-G}G^{\mu \nu }\bar{S}\gamma _{\mu }\partial _{\nu }S-\frac{1}{4}%
\sqrt{-G}G^{\mu \nu }G^{\sigma \lambda }{\bf F}_{\mu \sigma }{\bf F}_{\nu
\lambda }  \label{Lagf} \\
&&+\frac{i}{4}\varepsilon ^{\mu \nu \lambda }{\bf F}_{\mu \nu }\bar{S}\gamma
_{9}\psi _{\lambda }-\frac{1}{8}\sqrt{-G}G^{\mu \nu }\bar{S}\gamma _{9}\psi
_{\mu }\bar{S}\gamma _{9}\psi _{\nu },  \nonumber
\end{eqnarray}
where 
\begin{eqnarray*}
\,{\bf F}\,_{\mu \nu } &=&F_{\mu \nu }+i\bar{S}\gamma _{9}\gamma _{[\mu
}\psi _{\nu ]} \\
\bar{S} &=&S^{T}\gamma ^{0}=-S^{T}\frac{\gamma }{g},\,\,\, \partial _{\mu
}=\left( \partial _{0},\partial _{i}\right)
\end{eqnarray*}
and we used the following relations: 
\begin{eqnarray*}
\gamma _{j}\varepsilon ^{ji} &=&\gamma \gamma _{k}g^{ki}=-g^{ik}\gamma
_{k}\gamma ,\,\,\,\,\varepsilon _{ij}=g_{ik}g_{jl}\varepsilon
^{kl},\,\,\,\,\varepsilon ^{ik}\epsilon _{kj}=-\delta _{j}^{i} \\
\gamma _{i}\gamma _{j} &=&g_{ij}+\epsilon _{ij}\gamma ,\,\,\,\,\epsilon
_{ij}=\frac{\varepsilon _{ij}}{g},\,\,\,\, g_{ik}\varepsilon ^{kj}=\gamma
_{i}\gamma _{k}\varepsilon ^{kj}+\gamma \delta _{i}^{j}.
\end{eqnarray*}

We have thus shown that the effective action for the light--cone 10d super
D2-brane can be expressed in the form of a usual three-dimensional Abelian
gauge field coupled to the matter fields $X^{a},S$ in the induced
supergravity $\left( \ref{Met2}\right) $,$\left( \ref{grav}\right) $ defined
by the same matter fields - target space coordinates $X^{a},S$. The
expression for the zero component of the gravitino field $\left( \ref{grav}%
\right) $ is in accordance with the supersymmetry transformation of our
induced metric, which can be derived from $\left( \ref{susy1}\right) ,\left( 
\ref{Met2}\right) ,\left( \ref{3dgamma}\right) $ and $\left( \ref{grav}%
\right) $: 
\begin{equation}
\delta G_{\mu \nu }=i(\bar{\varepsilon }\gamma _{\mu }\psi _{\nu }+\bar{%
\varepsilon }\gamma _{\nu }\psi _{\mu }), \bar{\varepsilon }=\varepsilon ^{T}
\label{metsusy}
\end{equation}
So this transformation looks like some global remainder of the initial local
supersymmetry with the following choice of local supersymmetry parameter: 
\begin{eqnarray*}
\varepsilon (\tau ,\sigma _{i}) &=&-\gamma(\tau ,\sigma _{i}) \varepsilon \\
\gamma(\tau ,\sigma _{i}) &=&\frac{1}{2}\varepsilon ^{ij}\partial
_{i}X^{a}(\tau ,\sigma _{i})\partial _{j}X^{b}(\tau ,\sigma _{i})\frac{1}{2}%
\left[ \gamma ^{a},\gamma ^{b}\right]
\end{eqnarray*}
One should note that we have three $16\times 16$ induced curved space gamma
matrices, and our induced gravitino has $16$ spinor components. We therefore
have the $N=8$, $d=3$ supergravity multiplet \cite{Marcus} but our internal
and spinor spaces have mixed in the one reducible representation.

Finally we note that our covariant Lagrangian $\left( \ref{Lagf}\right) $ is
quadratic in the Maxwell field and therefore, as in the bosonic case, it is
possible to establish a direct duality relation with the M-theory membrane
action $\left( \ref{MT}\right) $.

For this reason we rewrite the membrane light--cone action $\left( \ref{MT}%
\right) $ in the covariant form: 
\begin{eqnarray}
L_{M} &=&-\frac{1}{2}\sqrt{-\tilde{G}}\tilde{G}^{\mu \nu }\partial _{\mu }X^{%
\dot{a}}\partial _{\nu }X^{\dot{a}}+\frac{1}{2}\sqrt{-\tilde{G}}  \label{MT1}
\\
&&+i\sqrt{-\tilde{G}}\tilde{G}^{\mu \nu }\bar{S}\tilde{\gamma}_{\mu
}\partial _{\nu }S,\,\,\,\,\, \bar{S}=S^{T}\tilde{\gamma}^{0}  \nonumber
\end{eqnarray}
Here the tilde metric and gamma matrices are induced from $9$ dimensional
transverse target space. Then we can separate in $\left( \ref{MT1}\right) $
the $X^{\dot{a}}$ into $8$- dimensional target space coordinates $X^{a}$ and 
$X^{9}$, and we can replace $\partial _{\mu }X^{9}$ by the independent
vector field $B_{\mu }$ and add to $L_{M}$ a metric independent topological
term \footnote{%
One may note that this topological term is supersymmetric because $\delta
B_{\mu }\sim \partial _{\mu }(\bar{\varepsilon}\gamma ^{9}S)$.}: 
\begin{equation}
S_{M}=\int L_{M}\left( X^{a},B\right) d\tau d^{2}\sigma +\frac{1}{2}\int BdA
\label{GenLag}
\end{equation}
With this and using our definition of the induced gravitino field $\left( 
\ref{grav}\right) $ and the following identity:

\[
\varepsilon ^{\mu \nu \lambda }\partial _{\mu }X^{9}\bar{S}\gamma _{9}\gamma
_{\nu }\psi _{\lambda }+\sqrt{-G}\partial _{\mu }X^{9}\bar{S}\gamma _{9}\psi
^{\mu }=-2\partial _{i}X^{9}S^{T}\gamma _{9}\psi _{j}\varepsilon ^{ij}, 
\]
we can rewrite $\left( \ref{GenLag}\right) $ in the following form: 
\begin{eqnarray}
S_{M}&&=\int d\tau d^{2}\sigma \left\{ -\frac{1}{2}\sqrt{-G}G^{\mu \nu }
\partial _{\mu }X^{a}\partial _{\nu}X^{a}+\frac{1}{2}\sqrt{-G}\right.  \nonumber \\
&&+i\sqrt{-G}G^{\mu \nu }\left( \bar{S}\gamma _{\mu }\partial _{\nu }S+
\frac{1}{2}B_{\mu }\bar{S}\gamma _{9}\psi _{\nu }\right)  \label{GenLag2} \\
&&\left.-\frac{1}{2}\sqrt{-G}G^{\mu \nu }B_{\mu }B_{\nu }+\frac{1}{2}\varepsilon
^{\mu \nu \lambda }B_{\mu }{\bf F}_{\nu \lambda }\right\}  \nonumber
\end{eqnarray}
The equation of motion for $A_{\mu }$ leads to $\left( \ref{MT1}\right) $.
But elimination of $B_{\mu }$ brings us back to our light-cone effective
Lagrangian for the super D2-brane $\left( \ref{Lagf}\right)$, which we
obtained from the Hamiltonian formulation. As mentioned above, the
connection between the 10d D2-brane and the 11d membrane was established and
exploited in \cite{Duff}, \cite{THS}, \cite{Sch}.

\section{Conclusions}

In the above the light--cone formalism has been developed for the
10--dimensio-nal super D2-brane, and it was shown, that all corresponding
equations of motion and constraints can be derived from the Lagrangian of a
3d Maxwell theory, interacting with matter fields, in a curved space-time
with a special induced 3d supergravity multiplet. This theory is invariant
with respect to the usual Abelian gauge transformations of gauge fields, and
with respect to area-preserving diffeomorphisms and supersymmetry. So, we
have shown, that as in the bosonic case \cite{MMM}, the non-linear super
D2-brane Lagrangian can be replaced in the light--cone gauge, by a quadratic
one over gauge fields, although the dependence on coordinates of the
membrane remains highly non-linear. The exact integration over gauge fields
can now be carried out, at least formally, and the corresponding determinant
has to be considered as an effective action for the super D2-brane, and can
be expanded in terms of the curvature tensor constructed from the induced
metric $\left( \ref{Met2}\right) $. Another direction in which to extend the
results of this article is to explore the connection between the light--cone
formulation with three dimensional extended supergravity and consideration
of the super Dp-brane in the light--cone formulation for p $>$ 2. The
connection between the D-brane actions and supergravity in the corresponding
dimensions is a long--standing and interesting problem. For p$=$2 the
connetion between Green--Schwarz and Neveu--Schwarz actions is well--knwon and
goes through the light--cone gauges. For higher dimensions some insight is
considered in the recent works \cite{kallosh}. Present consideration shows, for
the D2-brane, a possibility of introduction of the (induced) supergravity multiplet,
with the (induced) susy transformations. This is a field for future
investigations.

{\bf Acknowledgments}

This work was supported in part by the U.S. Civilian Research and
Development Foundation under Award \# 96-RP1-253 and by INTAS grants \#
96-538. R. Manvelyan is indebted to the A. von Humboldt Foundation for
financial support.

\end{document}